\documentstyle[psfig]{mn}

\begin{document}
\title[A double-frequency dwarf nova oscillation]{A double-frequency dwarf
nova oscillation in OY~Car}

\author[T. R. Marsh, Keith Horne]
{T. R. Marsh$^1$ and Keith Horne$^2$\\
$^1$Department of Physics and Astronomy, University of Southampton, 
Highfield, Southampton SO17 1BJ\\
$^2$School of Physics and Astronomy, University of St.\ Andrews,
North  Haugh, St.\ Andrews, Fife KY16 9SS}
\date{Accepted 28 May 1998
      Received ??
      in original form ??}
\pubyear{1998}

\maketitle

\begin{abstract}
We have detected coherent oscillations (``dwarf nova oscillations'')
in {\it Hubble Space Telescope} spectra of the dwarf nova OY~Car. The
oscillations were seen towards the end of a superoutburst of
OY~Car. The oscillations are extraordinary compared to the many other
examples in the literature for two reasons. First, their amplitude is
large, with a peak-to-peak variation of 8 to 20\% of the total flux
over the range 1100 to 2500\AA. However, most remarkably we find that
there are {\em two} components present simultaneously. Both have
periods close to $18\,{\rm s}$ (equivalent to $4800$ cycles/day) but
they are separated by $57.7\pm0.5$ cycles/day.  The lower frequency
component of the pair has a strong second harmonic while its
companion, which has about twice its amplitude, does not. The
oscillation spectra appear hotter than the mean spectrum and
approximately follow the continuum distribution of a black-body with a
temperature in the range $30$,$000$ to $50$,$000\,{\rm K}$.

We tentatively suggest that the weaker non-sinusoidal component
could represent the rotation of the white dwarf, although we have been
unable to recover any such signal in quiescent data.
\end{abstract}

\begin{keywords}
accretion, accretion discs -- novae, cataclysmic variables -- stars:
oscillations -- stars: individual: OY~Car
\end{keywords}

\section{Introduction}
Dwarf nova oscillations (DNOs) are one of the unsolved mysteries of
cataclysmic variable stars. First discovered by Warner \& Robinson
(1972), DNOs appear during outbursts as moderately coherent
oscillations (with $Q = P/\delta P$ from $10^4$ to $10^6$) with periods
from around 7 to 40 seconds, (although recently the lower limit has
been extended to $2.8$ seconds from observations of SS~Cygni, van
Teeseling 1997, Mauche 1997). The short periods clearly implicate the inner
accretion disc and white dwarf. However while the rotation of the
white dwarf is too coherent to match period changes seen in DNOs, it
is hard to see how the accretion disc can produce anything as coherent
as observed.  The amplitudes (peak-to-peak) of DNOs in optical
observations are typically less than $0.5$\% and therefore they can
often only be detected after period analysis. The amplitudes increase
towards shorter wavelengths and are tens of percents in X-ray light
curves (Cordova et al.\ 1984). The periods of DNOs correlate with
system brightness, becoming shorter as the system becomes brighter and
vice versa.

DNOs seen so far have exhibited a single, sinusoidal signal (although
quasi-periodic oscillations are occasionally seen as well). In this
paper we present data that violates both generalisations. We will show
that DNOs present in {\it Hubble Space Telescope} observations of the
short-period ($P = 91$ min) eclipsing dwarf-nova OY~Car have two
components, one of which has a strong second harmonic. We begin by
describing the observations.

\section{Observations}
\label{sec:obs}
On 24 April 1992 (pre-COSTAR) we took 419 spectra covering the range
1150 to 2510\AA\ with the G160L grating and the Faint Object
Spectrograph (FOS) on the Hubble Space Telescope.  Each exposure was
$4.74\,{\rm s}$ long, and the time from the start of one exposure to
the start of the next was $5.60\,{\rm s}$.

With the G160L grating, FOS also captures the zeroth order undispersed
light which is weighted to longer wavelengths.  The pass-band of the
zeroth order light has been calibrated by Eracleous et
al.\ (1994) who found that it has a full width at half maximum of
1900\AA\ centred on 3400\AA\ and who determined a scale factor of 820
counts/sec/mJy for our $4.3$'' aperture.

\label{sec:res}
\begin{figure}
\hspace*{\fill}
\psfig{file=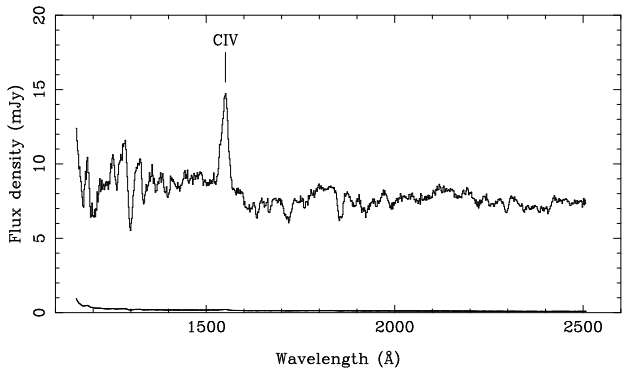,width=82mm}
\hspace*{\fill}
\caption{The mean spectrum of OY~Car observed at the end of a
superoutburst with the G160L grating and HST/FOS. The lower line
shows the $1\sigma$ uncertainties on the spectrum.}
\label{fig:avspec}
\end{figure}
\begin{figure*}
\hspace*{\fill}
\psfig{file=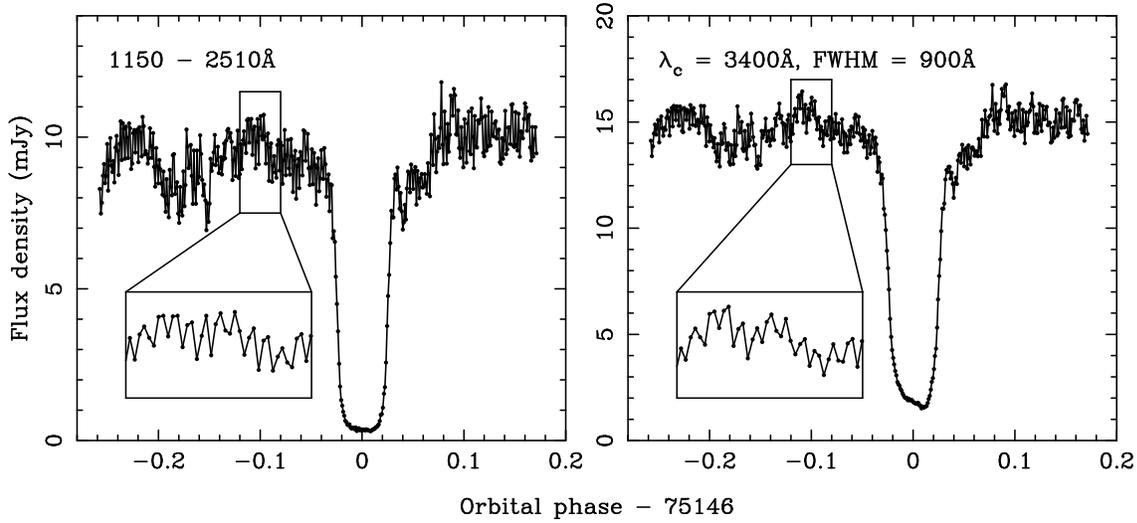,width=150mm}
\hspace*{\fill}
\caption{The light curve of OY~Car in the first order dispersed
light is plotted in the left panel whilst the right panel shows the
zeroth order light curve. Both light curves show periodic
oscillations. The oscillation is relatively stronger in the shorter
wavelength (left-hand) panel.}
\label{fig:lc}
\end{figure*}

OY~Car at this time was right at the end of a super-outburst which
started some 17 days earlier. All amateur measurements taken after our
run were upper limits, and so we believe that our data were taken
within a day of the return to quiescence.  Eight other {\it HST}
observations of the outburst were made prior to those we describe
here, the closest being observed two days earlier.  No oscillations
were found in any of these. Similarly, we find no oscillations in any
of 7 succeeding observations during quiescence, the first of which
took place 10 days after the observations reported here.

\section{Results}

The mean spectrum during the observations is presented in
Fig.~\ref{fig:avspec}. The spectrum is blue (note that it is plotted
in terms of $f_\nu$) with modest CIV emission. The continuum is very
complex with many features from what has previously been recognised as
the ``iron curtain'' of material that partially absorbs light from
the inner disc and white dwarf (Horne et al.\ 1994). The broad dips
around 1600 to 1800\AA\ and beyond 2200\AA\ are highly characteristic
in this regard.

It is the light curves (Fig.~\ref{fig:lc}) that prove to be unusual. 
As well as very deep eclipses characteristic of a small light source
concentrated at the centre of the disc, they exhibit what appears at
first to be a curious noise pattern. This is in fact the dwarf nova
oscillation, and is a rare case in which it can be seen directly
in the raw data. The mean flux level of around $9\,{\rm mJy}$ compares
to a level of about $1\,{\rm mJy}$ observed during quiescence.

\begin{figure}
\hspace*{\fill}
\psfig{file=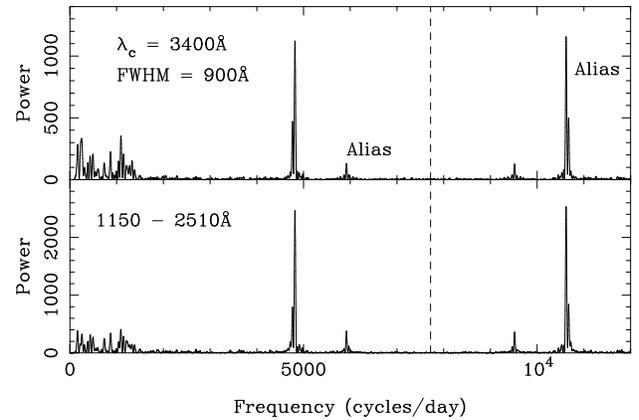,width=82mm}
\hspace*{\fill}
\caption{The Lomb-Scargle periodograms of the two orders are plotted
beyond the Nyquist frequency (dashed line). The peak at 5900
cycles/day is an alias of the second harmonic 
(seen near 9500 cycles/day) of the component near 4800 cycles/day.}
\label{fig:allperg}
\end{figure}

The periodogram (Lomb 1976; Scargle 1982) of each light curve is shown in
Fig.~\ref{fig:allperg}. 
These were computed after masking out data
affected by the eclipse 
and then fitting and subtracting a 7th order polynomial from the data
to remove long timescale variations. At high frequencies, the
periodograms are dominated by two peaks close to 4800 and 5900
cycles/day respectively. We will show below that the lower frequency
``peak'' is composed of two closely spaced signals. After some time we
realised that the higher frequency peak is the alias of a peak near
9500 cycles/day as we show in Fig.~\ref{fig:allperg} by extending it
beyond the Nyquist frequency (7719 cycles/day). We make this
identification because, as we will show, the 9500 cycles/day peak is
then the second harmonic of the {\em weaker and lower frequency}
component of the 4800 peak. The aliasing is a consequence of the
length of our exposures, which have the additional effect of reducing
the amplitude of the signal. If a signal of period $P$ is sampled with
exposures of length $\Delta t$, then the observed amplitude is $\sin
(\theta)/\theta$ times the true amplitude where $\theta = \pi \Delta
t/P$. In our case $\Delta t = 4.74\,{\rm s}$, whereas 9500 cycles/day
corresponds to $P = 9.09\,{\rm s}$, leading to a reduction factor of
$0.61$. In the periodogram, this is squared and means that the peak at
6000 cycles/day is only 37\%\ of its true height. For the same reason,
the third harmonic is reduced to only 7\%\ of its true
height. Unfortunately the aliasing means that the third harmonic is
expected to be at $1160$ cycles/day in a region of high
background from low frequency power, and it cannot be detected. 

%
%

\begin{figure}
\hspace*{\fill}
\psfig{file=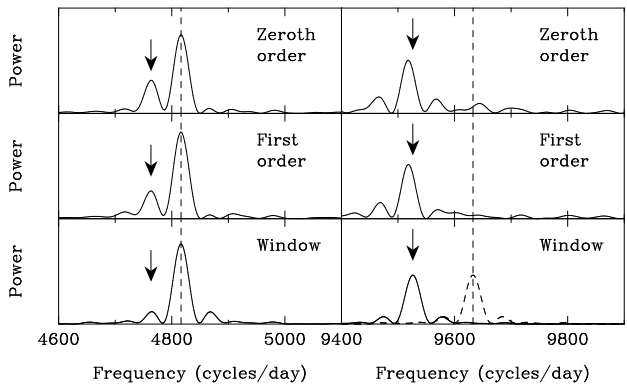,width=82mm}
\hspace*{\fill}
\caption{The figure shows regions around the two main peaks of the
periodogram of Fig.~\protect\ref{fig:allperg}. In the left-hand panels
the peak marked with an arrow is stronger in the data than in the
window function (generated by replacing the data with a pure
sinusoid). On the right, the arrows indicate the expected position of
the second harmonic of the peak marked by an arrow in the left-hand
panels. There is no sign of a second harmonic of the stronger peak
of the left-hand panel.}
\label{fig:zoomperg}
\end{figure}

We said above that the main peak has two components the weaker of
which has a second harmonic. This is shown in Fig.~\ref{fig:zoomperg}
where we isolate the two peaks at 4800 cycles/day and the second
harmonic at 9500 cycles/day. We also display ``window'' functions
computed by replacing the data by pure sinusoids, including the mask
around eclipse. In the left-hand series of panels, the data show a
peak immediately to the left of the main peak which is not seen as
strongly in the window function. The strongest peak is at $4816.32$
cycles/day while its companion is some $53$ cycles/day lower at
$4763.12$.  Unfortunately because of the short length of our
observation, there is a weak peak at the same place in the window
function and on this basis alone we would be hesitant in claiming the
presence of two signals. However in the right-hand panels, we see
the second harmonic of the $4763$ component while there is no
detectable second harmonic of its stronger companion. This shows that
the component  indicated by an arrow in the left-hand panels of
Fig.~\ref{fig:zoomperg} is not merely an artefact of sampling.

The Lomb-Scargle periodogram is not really appropriate in cases where
more than one frequency is present, although if they are far enough
apart we would not expect significant problems in determining the
frequencies of the peaks. In this case however, two frequencies are
close, and there must be doubt over the accuracy of the numbers quoted
above.  On the other hand, we need to know them accurately in order to
verify quantitatively our identification of the second harmonic. That
this is a significant problem is seen once uncertainties are placed on
the frequencies. Using Bayesian probability theory (as outlined in the
appendix) it is straightforward to extend the periodogram to account
for any number of periodic components. For a single component, this
reduces to a form very close to the Lomb-Scargle version, and shows
that the periodogram is closely related to the natural log of the
posterior probability distribution of the period. The curvature then
gives the uncertainty, and applying this we obtain the frequencies of
the three main peaks listed in the second column of
Table~\ref{tab:freq}, which can be compared to the values obtained
from the Lomb-Scargle periodogram listed in the first column.
\begin{table}
\centering
\begin{minipage}{83mm}
\caption{Frequencies of components}
\label{tab:freq}
\begin{tabular}{lccc}
Lomb- & \multicolumn{3}{c}{Bayesian models} \\
Scargle\footnote{The frequencies are all measured in units of
cycles/day} & 1 cpt            & 3 cpt           & 2 cpt ($f_3 = 2f_2$) \\[1mm]
$4816.32$ & $4816.26\pm0.27$ &$4816.59\pm0.34$ &$4816.72\pm0.31$ \\
$4763.12$ & $4763.19\pm0.42$ &$4760.13\pm1.21$ &$4759.05\pm0.33$ \\
$9518.41$ & $9518.36\pm0.65$ &$9517.94\pm0.68$ &$9518.10\pm0.66$ 
\end{tabular}
\end{minipage}
\end{table}
As expected the values in the first two columns of
table~\ref{tab:freq} are almost identical, but now with uncertainties
we can compare the frequencies of the peaks at $f_2 = 4763$ and $f_3 =
9518$ cycles/day that we claim are harmonically related to each
other. We find the difference $2f_2-f_3 = 8.02 \pm 1.06$ cycles/day,
and it would seem that there is a problem. This is the reason why in
Fig.~\ref{fig:zoomperg} the predicted second harmonic indicated by the
arrow in the right-hand panels is not a perfect match to the observed
peak.  However, once three components are included correctly (see the
appendix), we obtain the values in the third column. The frequencies
do indeed change, and now the difference between the harmonically
related components becomes $2f_2-f_3 = 2.32 \pm 1.39$, in acceptable
agreement within the uncertainties. Given the harmonic relation, we
can re-fit the frequencies on the basis that $f_3 = 2f_2$ exactly, so
that there are only 2 independent frequencies, and we arrive at the
fourth column which contains the final values that we will use from
now on. It can be seen that it is the harmonic at 9518 cycles/day which
dominates the determination of the frequency, and as we will show, it
has a higher amplitude than the fundamental.

The referee has pointed out that a single component with a phase
and/or period change might appear as more than one period in
periodograms, and that just such changes have been observed in other
examples of DNOs. This is certainly true. We carried out some tests
with a single signal of constant period suffering a sharp phase shift
about half way through the observations.  As the phase shift
increases, one of the two side peaks grows in strength while the main
peak shifts in frequency and weakens. When the phase shift reaches
about $120^\circ$ the periodogram appears not unlike
Fig.~\ref{fig:zoomperg}. In this case the ``true'' frequency is placed
somewhere between the two peaks, but is closest to the strongest
peak. The coincidence of the weaker peak (the $f_2$ component) with
one of the two side-lobes of the $f_1$ component is suspicious in this
regard. We reject this possibility however because of the $f_3$
component which agrees to within $1.7\sigma$ of $2f_2$ and yet is
$120\sigma$ away from $2f_1$. This means that $f_3$ is the
second harmonic of $f_2$, and thus component $f_2$ is genuinely
independent of $f_1$. We further note that while we know of no other
multiply periodic DNOs, multiple periodicity has been seen in the
dwarf nova WZ~Sge in quiescence (Patterson et al.\ 1998).

\subsection{Spectra of the oscillations}
%
%
%
%
%
%
%
%
%
%
%
%
Armed with accurate frequencies for the three components (two of them
harmonically related), we can fit sinusoids at many different
wavelengths to obtain the spectra of the oscillations. To do so we fit
all three components simultaneously to the spectra, using the form
$a\cos 2\pi f t + b \sin 2\pi f t$ for each component. This can be
re-expressed as $A \cos 2\pi (f t + \phi)$ where the semi-amplitude $A
= \sqrt{a^2 + b^2}$ and the phase $\phi$ is given by $\tan 2\pi\phi =
-b/a$. The estimate of amplitude $A$ suffers from a noise bias (it can
only be positive). Although this can be corrected approximately, it is
preferable to see first if there is any evidence for a variation of
phase with wavelength. If not, then $\phi$ can be fixed and we can
derive an unbiassed estimate of the semi-amplitude. The phase versus
wavelength
\begin{figure}
\hspace*{\fill}
\psfig{file=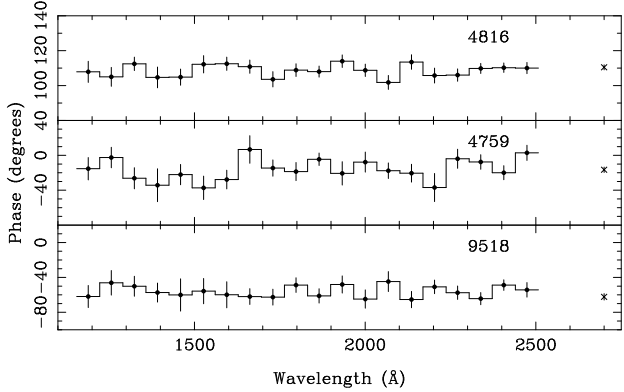,width=82mm}
\hspace*{\fill}
\caption{The phases of the 3 components of the oscillations are 
plotted versus wavelength. The right-most point is the phase of the
zeroth order light curve. There is no evidence for any dependence of
phase upon wavelength. The phases are measured with respect to the
zero point $HJD = 2448736.835$.}
\label{fig:oscphs}
\end{figure}
is plotted in Fig.~\ref{fig:oscphs} and shows that the phase is indeed
independent of wavelength in all three components. The weighted mean
phases of both orders was then computed in order to fit the
amplitudes. We obtain the values $110.6^\circ\pm0.8$,
$-15.4^\circ\pm1.9$ and $-57.7^\circ\pm1.9$ for the $4816$, $4759$ and
$9518$ components respectively. The latter two values show that the
$9518$ harmonic peaks $13.9^\circ\pm2.1$ after the $4759$ fundamental,
as measured in terms of the fundamental's phase. 

\begin{figure}
\hspace*{\fill}
\psfig{file=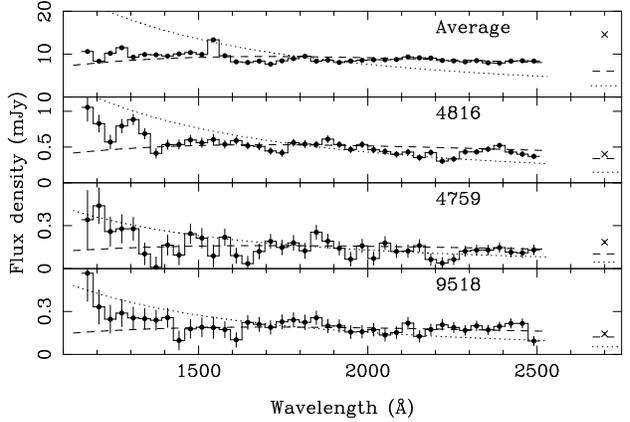,width=82mm}
\hspace*{\fill}
\caption{The semi-amplitudes of the 3 components of the oscillations are 
plotted versus wavelength along with the mean spectrum binned to the
same scale. The right-most point is the zeroth order point scaled
according to the calibration of Eracleous et al.\ (1994) and shifted in
wavelength to avoid compressing the first order spectrum; while the
statistical uncertainties on these points are low there is a 50\%
uncertainty in their calibration which would move them all up or 
down by the same factor.  The dashed lines show black-body spectra for
$T = 30$,$000\,{\rm K}$, while the dotted lines are Raleigh-Jeans
spectra. The values plotted are not corrected for the
amplitude reduction caused by the finite length exposures.}
\label{fig:oscamp}
\end{figure}

Holding the phases fixed we then fitted the amplitudes which are
plotted Fig.~\ref{fig:oscamp}.
The oscillation spectra show no sign of the CIV emission seen in the
mean spectrum (at 1550\AA). The CIV is thought to form by scattering
of continuum photons in an outflow above and below the disc. It
therefore gives us an angle-integrated view of the inner disc 
brightness. This suggests that the oscillations are not produced by 
the entire luminosity of the system varying, but rather by the varying 
visibility of some hot source. This counts against radial oscillation
models such as those of Molteni et al.\ (1996).

The oscillation spectra are very much hotter than the average
spectrum. This can be seen over the range of the first order dispersed
light but more clearly still in the ratio of first to zeroth order
light. This agrees with earlier work at optical wavelengths
(e.g. Middleditch \& Cordova 1982; Stiening et al.\ 1984); our data
extend this result into the space ultraviolet. Partly as a result, the
peak-to-peak amplitude in the first order light is very high, reaching
almost 20\% at the shortest wavelengths of the 4816 component. In the
zeroth order light, this has fallen to 5\%. Schoembs (1986) observed a
maximum amplitude of $2.2$\%, but given that his observations were
approximately B band (our zeroth order has a central wavelength of
$3400$\AA) and the very blue spectra, our values are probably
comparable.

The spectra of DNO are so blue that ground-based data can
only put a lower limit upon their temperature. With our coverage of 
the ultraviolet we might hope to fare better. Indeed, the
Raleigh-Jeans spectra plotted in Fig.~\ref{fig:oscamp} are
hotter than any of the oscillation spectra, particularly
in the small amount of zeroth order flux they produce. 
Allowing for the 50\% uncertainty Eracleous et al.\ (1994) quote on
their calibration of the zeroth order flux, we deduce an upper limit
of about $50$,$000\,{\rm K}$ if the oscillation spectra are
black-bodies. The lower limit is independent of the zeroth order
calibration since it is fixed by the short wavelength ultraviolet flux.
We place a lower limit of about $30$,$000\,{\rm K}$ (see 
Fig.~\ref{fig:oscamp}) from such a comparison.

\begin{figure}
\hspace*{\fill}
\psfig{file=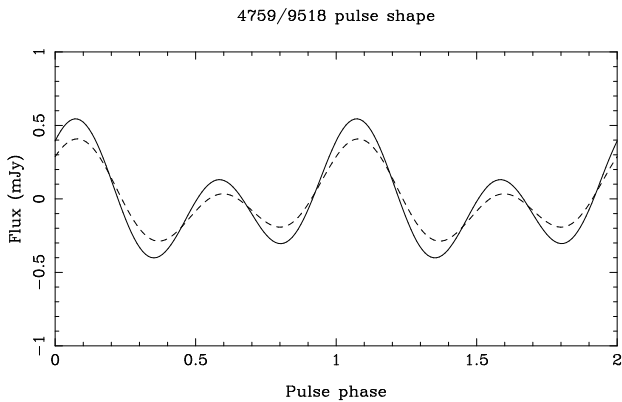,width=82mm}
\hspace*{\fill}
\caption{The shape of the 4759/9518 obtained from sinusoid fits and
corrected for the amplitude reduction caused by the finite exposure
lengths. The solid and dashed lines represent the first and zeroth
orders respectively. The pulse phase is measured in terms of the 4759
fundamental with respect to the zero point $HJD = 2448736.835$.}
\label{fig:pulse}
\end{figure}

The ratios of the oscillation amplitude in the first order light
divided by its amplitude in zeroth order (both measured in counts/sec)
are $1.37\pm0.04$, $1.10\pm0.08$ and $1.53\pm0.11$ for the 4816, 4759
and 9518 components respectively; the ratio of the mean fluxes in each
order is $0.68$. Thus the second harmonic 9518 component is hotter
than its fundamental counterpart at 4759 cycles/day. Moreover, the
amplitude of the second harmonic is larger than the fundamental.
We find the ratios of the 9518 to 4759 amplitudes (corrected for the
finite exposure factor discussed earlier) to be $1.58 \pm 0.09$ in the
first order and $1.14\pm 0.09$ in the zeroth order. Along with the
phases mentioned earlier, these lead  to the pulse shapes shown in
Fig.~\ref{fig:pulse}.
This plot shows a significant ``interpulse'' which will be of
significance in the discussion section. The true pulse shape may be
more complex since our exposures were too long to detect any but the
first two harmonics.

\subsection{The eclipse of the oscillations}
\begin{figure*}
\hspace*{\fill}
\psfig{file=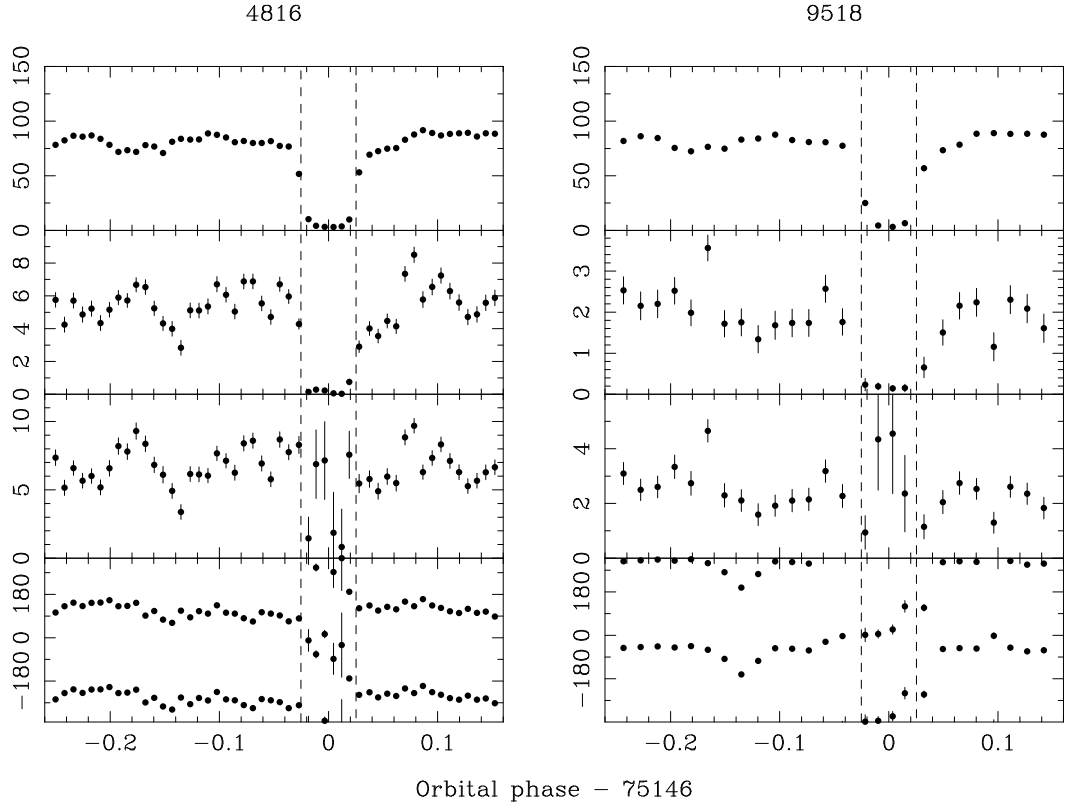,width=140mm}
\hspace*{\fill}
\caption{From the top are plotted the mean flux, oscillation
amplitude, percentage oscillation amplitude and oscillation phase 
(degrees) evaluated over data segments of 16 points for
the 4816 component and 30 for the 9518 component for the first order
light.}
\label{fig:uv_vs_time}
\end{figure*}
\begin{figure*}
\hspace*{\fill}
\psfig{file=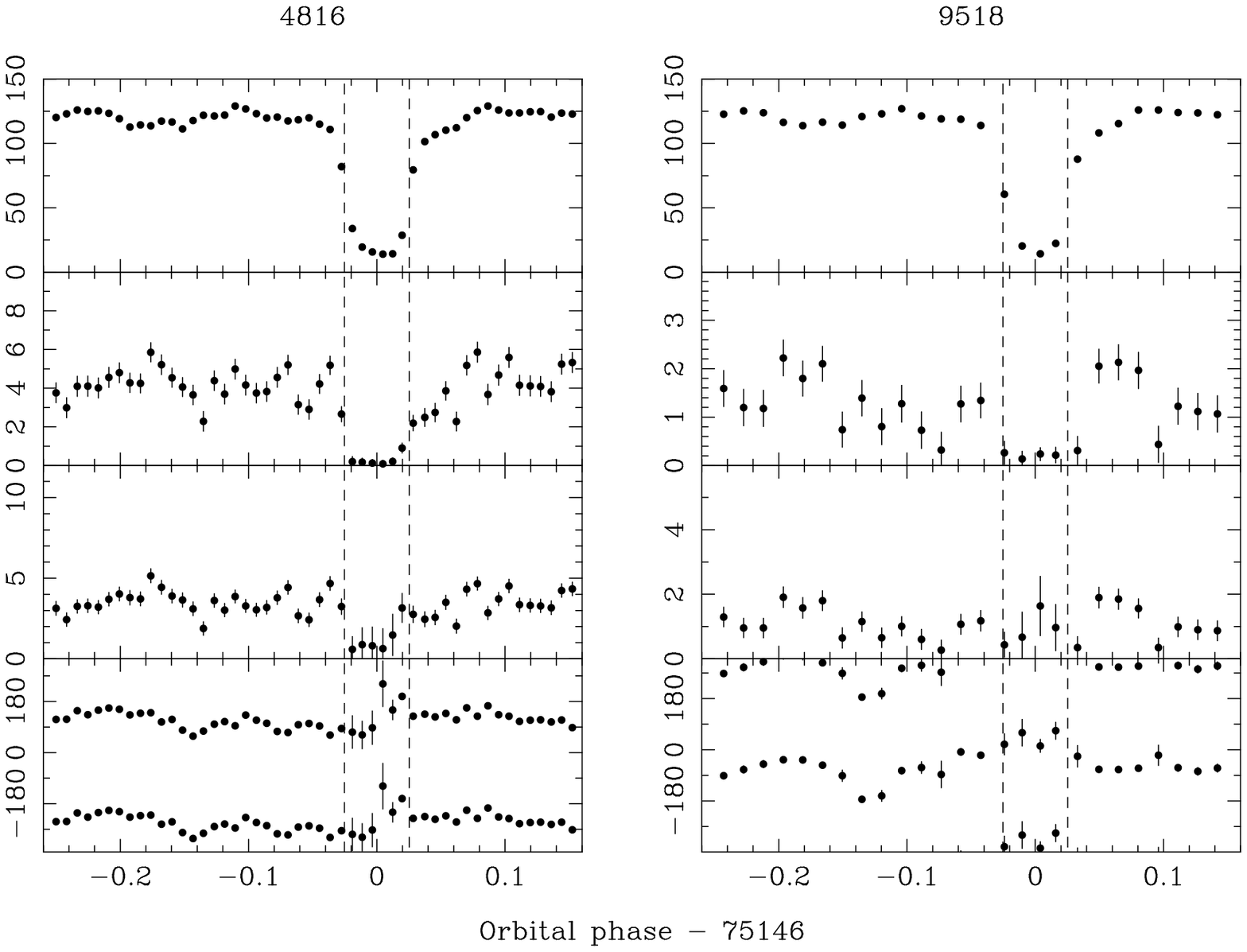,width=140mm}
\hspace*{\fill}
\caption{From the top are plotted the mean flux, oscillation
amplitude, percentage oscillation amplitude and oscillation phase 
(degrees) evaluated over data segments of 16 points for
the 4816 component and 30 for the 9518 component for the zeroth order
light.}
\label{fig:opt_vs_time}
\end{figure*}
As OY~Car is eclipsing, we can learn something about the distribution
of the oscillating light by measuring the amplitude and phase of the
oscillations as a function of time.  The standard method of doing so
is to hold the period fixed and then fit sinusoids to short sections
of the data. How short a section can be used depends upon the strength
of the oscillations. Our oscillations are especially strong, but we
have the extra complication of multiple periods. While it is almost as
easy to fit to several periods as it is to fit to one, the solutions become
ill-conditioned if data segments are not long enough to separate the
periods. This is most obviously a problem with the 4816 and 4759
components which can only barely be separated using all of our data.
Therefore for the purposes of this section we simply ignore the weaker
component (4759) while realising that it will cause a certain amount
of interference or ``beating''.
Unfortunately spectral leakage from the 4816 component can also affect
the 9518 component, which since it appears at 5900 cycles/day because
of aliasing is much closer than it seems. We applied a linear taper to
the data segments in an effort to reduce the leakage, but were forced
to use segments of 30 points each for the 9518 component. For the
4816/4759 combination we settled on 16 points per segment, and in each
case we used a 50\% overlap from one segment to the next. Prior to
splitting the data up we also applied a high-pass filter to remove low
frequencies in order to prevent leakage from them affecting the
results.  The results of the fits are shown in
Figs.~\ref{fig:uv_vs_time} and \ref{fig:opt_vs_time} for the first and
zeroth order light respectively.

In interpreting the plots it should be remembered that the amplitude
measurements are subject to a positive bias of the same order as the
uncertainties plotted and so the amplitudes during eclipse are
consistent with no detection even though it appears that with 
several marginal detections there is some evidence
for oscillations during eclipse. Both the mean flux and oscillations
are deeply eclipsed, but from the zeroth order results on the 4816
component in Fig.~\ref{fig:opt_vs_time} we can say that the
oscillations are even more deeply eclipsed than is the mean light.
The other panels give no clear answer, partly because
the eclipse is very deep even in the mean light, and partly because of
the noise bias discussed earlier. 

Since we know that the period of the oscillations imply an origin in
the inner disc or white dwarf, it may not seem that we have learned
much. However, in other systems, such as DQ~Her and UX~UMa, the
oscillating light is spread out, indicating that we see light
reprocessed from the disc. Phase shifts observed during eclipse show
where the reprocessing occurs (Petterson 1980). It seems then that
such effects play at best a minor role in our data; we will look at
reprocessing in more detail in section~\ref{sec:reproc}.

\section{Discussion}
Our {\it HST} data are qualitatively different from previous
observations of dwarf nova oscillations. The essential features of our
data are (a) the simultaneous presence of two oscillations each with a
period close to $18\,{\rm s}$ but separated by $57.7\pm0.5$
cycles/day, and (b) a strong second harmonic of the weaker and lower
frequency component which peaks just $13^\circ$ after the fundamental,
and no second harmonic of the stronger component.

To our knowledge, there are no previous cases of either multiple or
non-sinusoidal oscillations. As has long been realised, the short
periods of DNO imply an origin in or near the white dwarf. The direct
interpretation of our data is that this region is capable of producing
two oscillations of different character at the same time. Although it
is not clear why it has not been seen before, it is what we believe is
happening. However, before proceeding upon this basis, we will first
consider an attractive alternative which turns out not to be 
satisfactory. 

\subsection{Reprocessing models}
\label{sec:reproc}
We already know of one class of cataclysmic variable star in which a
signal at one period can give rise to others. These are the DQ~Her
stars in which a coherent signal from the rotation of a magnetic white
dwarf can be detected. These stars also show sidebands to the rotation
frequency displaced by the orbital frequency. Such sidebands can
result from modulation of the spin frequency $\omega$ on the orbital
frequency $\Omega$ which leads to $\omega\pm\Omega$ sidebands. A
single sideband at $\omega-\Omega$ can result from reprocessing of the
spin pulses off structures fixed in the rotating frame or from direct
interaction of the gas stream with the magnetosphere of the white
dwarf. It is clear then that such effects are possible, but could they
be occurring in OY~Car?

The DQ~Her analogy suggests that the $4759$ cycles/day component might
be a sideband of the 4816 component. We expect it to be this way round
because the 4816 component is the stronger and because if the light
source and reprocessing structure orbit in the same direction, the
reprocessed frequency is lower than the main signal. In addition the
4816 component is sinusoidal and thus can be identified with the
``standard'' DNO of other observations. The reprocessing structure
must advance in the inertial frame at the $57.7\pm0.5$ cycles/day
splitting we measured earlier. This is $3.6 \pm 0.03$ times the
orbital frequency in OY~Car. During superoutburst it is thought that
the accretion disc reaches a radius where it orbits 3 times for each
orbit of the binary (Whitehurst 1988). The 3:1 resonance then drives the
disc into an eccentric shape and subsequent precession can then
explain the observation of flares or ``superhumps'' on periods longer
than the binary period. From Kepler's third law, a ratio of $3.6$
corresponds to a radius of about $89$\% of the 3:1 radius. Since our
observations were placed right at the end of outburst when the disc
should have shrunk somewhat, this seems plausible. There are many
other constraints that such a model must satisfy however.

First of all there must be an asymmetry in the outer disc so that we
can see it lit up every time the beacon from the inner disc sweeps by
it. One cannot expect any such asymmetry to last very long since shear
in the disc always tends to smear out azimuthal structure. While we
cannot suggest a mechanism for generating such an asymmetry, its 
short-lived nature could be viewed as a plus point since then such
extra components would be rarely seen. In order to be consistent with
the eclipse results, we also have to suppose that the asymmetry is
small enough to be totally eclipsed. 

Even allowing the above constraints, there are higher hurdles for
reprocessing models to face.  The blue colours of the oscillation
spectra imply a high effective temperature, particularly so since our
spectra extend into the space ultraviolet. A comparison of the zeroth
to first order flux ratio with black-body models of reprocessing
suggest an effective temperature of at least $20$,$000\,{\rm K}$.
This is lower than the $30$,$000\,{\rm K}$ limit discussed earlier
because oscillation spectra can appear more blue than the source of
light that produces them since they depend upon the derivative of the
spectrum with respect to temperature. The luminosity $L$ needed to
produce reprocessed radiation with effective temperature $T_{\rm eff}$
at radius $R$ in the disc is given by
\[ L = 4\pi R^2 \sigma T^4_{\rm eff} \frac{d \ln R}{d (H/R)} ,\]
where $H$ is the scale height in the disc. Taking $H/R \propto
R^\beta$, then the derivative term becomes $1/(\beta H/R)$. Typically
$\beta \sim 1/8$ (Shakura \& Sunyaev 1973) and we take $H/R \sim 1/20$
and so the derivative term is of order $160$.

Taking Wood \&\ Horne's (1990) parameters for OY~Car, the $3.6$:1
radius corresponds to $R = 0.41 a$ or $1.9\times 10^8\,{\rm m}$.  We
then find that to reach our minimum of $20$,$000\,{\rm K}$ the central
peak luminosity must be of order $L = 1700\,{\rm L}_\odot$.  The
accretion luminosity of the inner disc and boundary layer,
$GM\dot{M}/R$, set an upper limit to the mean luminosity of the
oscillating source (which is half the peak luminosity assuming 100\%
modulation). We find $\dot{M} > 3 \times 10^{-7} \,{\rm M}_\odot/{\rm
yr}$, which is higher than the maximum rate expected for dwarf novae,
and a strong point against reprocessing. 

The second harmonic is perhaps even more difficult to understand on
the reprocessing model. It requires a non-linear repose
to the incident flux and can be generated if as the flux incident on
the disc rises, the reprocessing efficiency rises as well, thus
sharpening the peaks. However even if this process is taken to its
extreme to produce a series of sharp spikes or delta-functions, we
can only obtain a second harmonic equal to the fundamental. In our case
the second harmonic is stronger than the fundamental by a factor 
of $1.58 \pm 0.09$ in the first order light and $1.14\pm 0.09$ in
the zeroth order as illustrated in Fig.~\ref{fig:pulse}. In order
to obtain the pulse shapes seen one is forced to suppose a complex 
response in which the reprocessing efficiency at first decreases and 
then increases with rising incident flux. We find this very implausible.

Our conclusion is that a single oscillation plus reprocessing cannot
explain our observations: there really are two oscillations present
at the same time. 

\subsection{Two oscillations}
Accepting that both oscillations are generated on or near the white
dwarf, how do they originate? The trapped $g$- and $r$-mode model of
Papaloizou \& Pringle (1978) can certainly produce more than one
oscillation frequency. Indeed, it is hard to see how this model could
only produce one frequency, an argument that has been used in the past
against it. Now we have seen more than one component. However, in our
view this does not support Papaloizou \& Pringle's model because each
component has a very different character, and if two oscillations are
possible, why not more?

Another model of DNOs frequently discussed is one in which the DNO
frequency represents the keplerian orbital frequency of some part of
the inner disc (Bath 1973; Mauche 1996). On this model for instance,
the 4816 component might represent the keplerian frequency, and, if
so, it corresponds to an orbital radius of $1.4$ times the radius of
the white dwarf on the physical parameters of Wood \& Horne
(1990). Explaining the relatively high coherence of DNOs has always
seemed difficult for keplerian models as why should just one specific
orbital frequency dominate?  Our data perhaps provide some support
however because there is an appealing explanation for the 4759/9518
combination, which might represent either the white dwarf rotation
frequency or the beat frequency between the white dwarf and keplerian
frequencies. In the latter case the white dwarf would rotate at the
frequency difference of $57.7\pm0.5$ cycles/day. An attractive feature
of this model is that it is entirely natural that the two components
should have a different nature and second harmonics are easily
produced if the white dwarf has two accreting poles for instance.
Given a choice between 4759 or $57.7$ cycles/day for the rotation
frequency of the white dwarf, we would favour 4759 since we already
know of short rotation period examples such as AE~Aqr and DQ~Her,
amongst weakly magnetic accretors, a category which OY~Car surely
falls into. On the contrary, it is hard to imagine how the white dwarf
in OY~Car could have the $25$ minute period required to match the
$57.7$ frequency splitting as this would be longer than in many of the
longer period DQ~Her stars. On the basis of this idea, we searched for
coherent periodicities in 7 epochs of {\it HST} taken during
quiescence following the outburst and in Wood et al.'s (1989)
ground-based photometry. We failed to find any coherent signals, with
detection levels of $0.1$ and $0.3$\% respectively (these limits only
apply above 1000 cycles/day, because orbital variations and flickering
limit the sensitivity at low frequencies).  In this respect at least
the 4759/9518 component behaves like other DNOs which have never been
seen in quiescence.

It is not clear how much the failure to detect the 4759/9518 component
in quiescence counts against its being the rotation frequency of the
white dwarf. We speculate that perhaps it only becomes evident when
the inner disc orbital period (represented by the 4816 component) is
close to co-rotation with the white dwarf (4759 component).  Such a
situation may favour the largest difference between accretion rate
onto the poles compared with other parts of the white dwarf making the
white dwarf spin period detectable. We note that the shortest DNO period
from OY~Car observed by Schoembs (1986) of $19.44$ seconds is over $300$
cycles/day away from 4759 and may have already been too different to
generate the 4759/9518 signal.

A third, hybrid model has been put forward by Warner (1995), based
upon an earlier model of Paczynski (1978). In this model the DNO
frequency is set by the outer layers of a weakly magnetic white dwarf
decoupling from the white dwarf itself. The variation in period is put
down to the spin up/down of the layer with accretion rate.  This model
suffers from the same problem as the keplerian models in that it is
hard to see how a coherent oscillation can result. Now with our data,
there is an additional difficulty as Warner's model does not appear to
have any natural means for producing more than one periodicity.

In summary while none of the current models provide a convincing 
explanation for DNOs, our observation of two components seems more
naturally explained by keplerian orbit models than any other.

\section{Conclusions}
We have found dwarf nova oscillations at the end of a super-outburst
observed with {\it HST}. The oscillations have a large amplitude,
reaching 20\% peak to peak at the shortest ultraviolet wavelengths.
Most remarkably there are two oscillations present with similar
periods, and the weaker and longer period of the two has a strong
second harmonic.

We discuss several current models and while none are persuasive,
models in which the DNO frequency represents an orbital frequency of
the inner disc give the most natural explanation of the additional
component as it can represent the rotation of the white
dwarf. However, we have been unable to find any sign of such a
component during quiescence, and so, if our suggestion is true, some
means of suppressing the signal is required.

\section*{Acknowledgments}
TRM was supported by a PPARC Advanced Fellowship during the course of
this work.  The data reduction and analysis were carried out on the
Southampton node of the UK STARLINK computer network. We are grateful
to Janet Wood for making her ground-based photometry available to us.

\appendix
\section{Bayesian Period Measurement}
The detection and measurement of periodic components can be handled
using Bayesian probability theory (e.g.\ Gregory \& Loredo 1992;
Bretthorst, 1988). We consider the probability that our data is
consistent with zero (model $Z$) versus the probability that it can be
represented by a sum of $N$ sinusoids (model $S$). In practice, these
models imply that data must be ``de-trended'' by fitting polynomials
or splines to remove low frequency variability prior to analysis.

We wish to compute the probability ratio of these models given the data,
$P(S|D)/P(Z|D)$. Let the flux at time $t_i$ be 
$y_i$, with uncertainty $\sigma_i$, for $i = 1$ to $M$. Our
sinusoid can be represented in linear form as
\[ y'_i = \sum_{j=1}^{2N} x_{i,j} a_j ,\]
where the $x_{i,j}$ functions are alternately the value of cosine or
sine evaluated for frequency $j$ at $t_i$. Thus $x_{i,1} = \cos 2\pi
f_1 t_i$, $x_{i,2} = \sin 2\pi f_1 t_i$, $x_{i,3} = \cos 2\pi f_2t_i$, etc.
Written in this fashion rather than with phases makes later
integrations simpler.

Bayes' theorem gives
\[ \frac{P(S|D)}{P(Z|D)} = \frac{P(S)}{P(Z)} \frac{P(D|S)}{P(D|Z)} .\]
The first term on the right expresses one's prior belief in each model.
It is important when one is discussing whether a component is really
present, but since it remains constant, it does affect the periods
measured. Information about the periods is contained in the second
term, which is the ratio of the probability of our obtaining the data
given the models.

Let us first consider $P(D|Z)$. We assume that the data are gaussian
random variables then 
\[ P(D|Z) = \frac{1}{\Pi_{i=1}^M\sqrt{2\pi}\sigma_i} 
\exp \left\{ -\sum_{i=1}^M \frac{y_i^2}{2\sigma_i^2}\right\}.\]
The probability of the data given the sinusoid model is more
complicated because we first need to compute it for a specific vector
of parameters, i.e. $P(D|S,\bmath{a})$. This is straightforwardly
\[ P(D|S,\bmath{a}) = \frac{1}{\Pi_{i=1}^M \sqrt{2\pi}\sigma_i} 
\exp \left\{-\sum_{i=1}^M \frac{(y_i-y'_i)^2}{2\sigma_i^2}\right\}.\]

In order to obtain the probability we want, $P(D|S)$, we must
integrate over the parameters ${\bf a}$:
\[ P(D|S) = \int\!\int\!\int \cdots \int P(D|S,\bmath{a}) 
P(\bmath{a})\,d\bmath{a}. \]
Once again $P(\bmath{a})$ is a prior probability, in this case of
the semi-amplitudes. We take the amplitudes to have independent,
uniform distributions over the range $-R/2$ to $+R/2$ so
\[ P(\bmath{a}) = \frac{1}{R^{2N}}.\]
One normally has little idea of what $R$ would be, so it should be set
conservatively. For example, setting $R$ equal to the brightest OY~Car
has ever been seen would be a reasonable course to take. Its value has
some effect on the significance of a peak but not on the value of the
period deduced.

With these assumptions we now need to integrate the exponential over a
$2N$ dimensional cube of side $R$. To make this simpler we assume
that this cube contains all the region over which the exponential is
significant. After some algebra we obtain 
\begin{eqnarray*}
P(D|S) &=& \frac{1}{\Pi_{i=1}^M \sqrt{2\pi}\sigma_i} 
\exp \left\{-\sum_{i=1}^M \frac{y_i^2}{2\sigma_i^2}\right\} \\
& & \times \left(\frac{\sqrt{2\pi}}{R}\right)^{2N} 
\frac{1}{\sqrt{\det \bmath{X}}}
\exp \left\{\frac{1}{2} \bmath{b}^t \bmath{X}^{-1} \bmath{b}\right\},
\end{eqnarray*}
where the vector $\bmath{b}$ is given by
\[
\bmath{b} = \left( \begin{array}{c} 
\sum_i w_i x_{i,1} \\ 
\sum_i w_i x_{i,2} \\
\vdots \\
\sum_i w_i x_{i,2N}
\end{array} \right)
\]
and the $2N$ by $2N$ symmetric matrix $\bmath{X}$ is given by
\[
\bmath{X} = 
\left( \begin{array}{ccc} 
\sum_i w_i x_{i,1} x_{i,1} & \cdots & \sum_i w_i x_{i,1} x_{i,2N} \\
\sum_i w_i x_{i,2} x_{i,1} & \cdots & \sum_i w_i x_{i,2} x_{i,2N} \\
\vdots & & \vdots \\
\sum_i w_i x_{i,2N} x_{i,1} & \cdots & \sum_i w_i x_{i,2N} x_{i,2N} 
\end{array}
\right) 
\]
where $w_i = 1/\sigma_i^2$ and the sums extend from $i=1$ to $M$, the
number of data points.

Finally dividing by $P(D|Z)$ and taking the natural logarithm we
obtain
\begin{equation}
 \ln \frac{P(D|S)}{P(D|Z)} = 2N \ln \frac{\sqrt{2\pi}}{R} -
\frac{1}{2} \ln (\det \bmath{X}) + \frac{1}{2} 
\bmath{b}^t \bmath{X}^{-1} \bmath{b}. \label{eq:ratio}
\end{equation}
Together the first two factors represent the ratio of the volume of
parameter space allowed by the data compared to the volume ($R^{2N}$)
allowed by the prior model of the semi-amplitudes. This is a factor
penalising the sinusoid model for its adjustable parameters. The last
factor is the important one. For example, if we take a single sinusoid
model with uniform uncertainty values, one can show that the value of
$\bmath{b}^t \bmath{X}^{-1} \bmath{b}/2$ is identical to the
expression given by Scargle (1988) for a periodogram of unequally
spaced data.

The best frequencies are those which maximise the log probability
ratio of Eq.~\ref{eq:ratio} (which is why $R$ has no effect upon their
values). The maximum can be located using standard techniques. If at
the same time the second derivative matrix is evaluated, then
uncertainties can be computed. This is the procedure we applied to
derive the frequencies listed in table~\ref{tab:freq} of this paper.
\end{document}